\documentclass{aa}
\usepackage{graphicx}
\usepackage{txfonts}

\newcommand{\vertical}[1]{
   \begin{sideways}
      #1
   \end{sideways}
}
\begin{document} 

   \title{The long-term evolution of known resonant trans-Neptunian objects}

   \author{M. Saillenfest\inst{1,2} \and G. Lari\inst{2}}

   \institute{IMCCE, Observatoire de Paris, 77 av. Denfert-Rochereau, 75014 Paris, France\\
              \email{melaine.saillenfest@obspm.fr}
         \and
             Dipartimento di Matematica, Università di Pisa, Largo Bruno Pontecorvo 5, 56127 Pisa, Italia\\
             \email{lari@student.dm.unipi.it}
             }

   \date{Received January 30, 2017; accepted April 10, 2017}
 
   \abstract
  % context heading (optional)
   {}
  % aims heading (mandatory)
   {Numerous trans-Neptunian objects are known to be in mean-motion resonance with Neptune. We aim to describe their long-term orbital evolution (both past and future) by means of a one-degree-of-freedom secular model. In this paper, we focus only on objects with a semi-major axis larger than 50 astronomical units (au).}
  % methods heading (mandatory)
   {For each resonant object considered, a 500\,000-year numerical integration is performed. The output is digitally filtered to get the parameters of the resonant secular model. Their long-term (Giga-year) orbital evolution is then represented by the level curves of the secular Hamiltonian.}
  % results heading (mandatory)
   {For the majority of objects considered, the mean-motion resonance has little impact on the long-term trajectories (the secular dynamics is similar to a non-resonant one). However, a subset of objects is strongly affected by the resonance, producing moderately-high-amplitude oscillations of the perihelion distance and/or libration of the argument of perihelion around a fixed centre. Moreover, the high perihelion distance of the object 2015\,FJ345 is plainly explained by long-term resonant dynamics, allowing us to also deduce its orbital elements at the time of capture in resonance (at least 15 million years ago). The same type of past evolution is expected for 2014\,FZ71.}
  % conclusions heading (optional)
   {}

   \keywords{trans-Neptunian object -- mean-motion resonance -- secular model}

   \maketitle

\section{Introduction}
   The orbital distribution of the trans-Neptunian objects is still an open problem, leading to numerous conjectures about the origin and evolution of the external Solar System. In this context, mean-motion resonances with Neptune are known to have played an important role in sculpting this population from its initial distribution \citep[see for instance][]{MALHOTRA_1993,DUNCAN-etal_1995,GOMES-etal_2005}. According to numerical simulations, temporary and chaotic resonance captures are the most frequent, but some trappings are sustainable on a Giga-year timescale, in particular when the resonant link leads the small body towards a dynamical region where the semi-major axis cannot diffuse anymore. This happens for instance if the perihelion distance increases during the resonant motion, because this reduces the width of every neighbouring resonance, preventing the object from undergoing chaotic diffusion driven by resonance overlap. The resulting trajectory is quasi-integrable and it can be modelled by a secular-like theory.
   
   Recently, \citet{SAILLENFEST-etal_2016a} presented a semi-analytical model designed to accurately describe, and in a plain way, the long-term dynamics of such objects. This model can be seen as a generalisation of the secular theory by \citet{KOZAI_1985} and its extension by \citet{GALLARDO-etal_2012}. In order to reduce the system to a single degree of freedom, the key element of the resolution consists in the large separation of timescales between the oscillation of the resonant angle and the precession of the argument of perihelion (adiabatic approximation). Up to now, the application of this model has been made exclusively on distant small bodies ($a>100$~au), for which the adiabatic approximation is more relevant. However, only a few trans-Neptunian objects are known with semi-major axes that large, and their orbital uncertainties are necessarily rather high since only a fraction of their orbits has been observed (orbital period $>1000$ years). Consequently, the application of a secular model to these objects can only give the long-term dynamics these objects would have if they were in mean-motion resonance with Neptune \citep{SAILLENFEST-etal_2016b}.
   
   In this paper, our goal is to study the long-term evolution of the trans-Neptunian objects known to be in resonance, that is, those for which the uncertainties of the orbital elements are predominantly compatible with a resonant trapping. As presented for instance by \citet{LYKAWKA-MUKAI_2007}, these objects are quite numerous and most of them have small semi-major axes (because of the observational constraints). We show here that, fortunately, the adiabatic approximation is also viable for semi-major axes as small as $50$~au, even if the two timescales are not as strikingly separated as for $a>100$~au. For even closer objects, individual analyses would be required, not only to check the validity of the adiabatic approximation, but also the possible occurrence of secular resonances with the giant planets \citep{KNEZEVIC-etal_1991}, which would also invalidate this model. Consequently, this article focusses on the $\sim\!40$ currently known resonant trans-Neptunian objects with $a>50$~au.
   
   We selected the known resonant trans-Neptunian objects from three reference classifications \citep{LYKAWKA-MUKAI_2007,GLADMAN-etal_2008,GLADMAN-etal_2012}, retaining only the ``securely classified'' ones with a semi-major axis larger than $50$~au. Two recent observation reports were also added: \citet{BANNISTER-etal_2016} present the discovery of 2015\,RR$_{245}$, which has a confirmed resonant classification; on the other hand, \citet{SHEPPARD-etal_2016} introduce some objects that are probably resonant or have been strongly affected by a mean-motion resonance with Neptune. This last classification is slightly different, but we chose to add these objects to this study to emphasise the link with recent results focussed on high-perihelion objects. Indeed, the secular model used in this paper proved to be an efficient tool to reveal dynamical paths leading to high perihelion distances and large inclinations. The objects studied in this paper are listed in Tab.~\ref{tab:list}.
   
   In Sect.~\ref{sec:model}, we recall briefly the resonant secular model used (Sect.~\ref{ssec:model}) and we explain how suitable secular parameters can be obtained from known objects (Sect.~\ref{ssec:param}). Then, Sect.~\ref{sec:res} summarises our results using a selection of the most representative phase portraits obtained. Most of the objects considered have quite ordinary secular dynamics (Sect.~\ref{ssec:gen}), but a subset of them evolve near or inside secular libration islands (Sect.~\ref{ssec:libr}) and one object follows a regular-by-part secular trajectory (Sect.~\ref{ssec:FJ345}).
   
\section{Model and method}\label{sec:model}
   \subsection{The resonant secular model}\label{ssec:model}
   In this section, we outline the semi-analytical model introduced by \citet{SAILLENFEST-etal_2016a}. We consider a massless particle affected by the Sun and $N$ planets, which are supposed on circular and coplanar orbits (typically Jupiter, Saturn, Uranus and Neptune in their invariable plane). The dynamics of the particle are studied in the exterior mean-motion resonance $k_p\!:\!k$ with the planet $p$. We introduce the resonant angle:
   \begin{equation}\label{eq:sigma}
   \centering
      \sigma = k\,\lambda - k_p\,\lambda_p - (k-k_p)\,\varpi
   \end{equation}
   with $k,k_p\in\mathbb{N}$ and $k>k_p$. In this expression, $\lambda$ and $\varpi$ are the mean longitude and the longitude of perihelion of the small body, and $\lambda_p$ is the mean longitude of the planet $p$. From the Hamiltonian in osculating coordinates, we get the ``semi-secular'' dynamical system by a perturbative method to first order. The resulting Hamiltonian can be written as:
   \begin{equation}\label{eq:Kssec}
      \mathcal{K}\big(\Sigma,U,V,\sigma,u\big) = -\frac{\mu^2}{2\,(k\Sigma)^2} - n_pk_p\Sigma + \varepsilon\,\mathcal{K}_1\big(\Sigma,U,V,\sigma,u\big)
   \end{equation}
   where $\mu$ is the gravitational parameter of the Sun and $n_p$ is the mean motion of the planet $p$. In that expression, $\varepsilon\,\mathcal{K}_1$ is the numerically computed average of the perturbative part of the osculating Hamiltonian over the short-period angles. The canonical coordinates used can be written in terms of the usual heliocentric Keplerian elements $(a,e,I,\omega,\Omega,M)$ of the particle:
   \begin{equation}\label{eq:SUV}
      \left\{
      \begin{aligned}
         \Sigma &= \sqrt{\mu a}/k\\
         U &= \sqrt{\mu a} \left(\sqrt{1-e^2} - k_p/k\right) \\
         V &= \sqrt{\mu a} \left(\sqrt{1-e^2}\cos I - k_p/k\right)
      \end{aligned}
      \right.
   \end{equation}
   along with their conjugated angles $\sigma$ and $(u,v)=(\omega,\Omega)$. Since the Hamiltonian~\eqref{eq:Kssec} does not depend on $v$, the quantity $V$ is a semi-secular constant of motion. The pairs of coordinates $(\Sigma,\sigma)$ and $(U,u)$ evolve generally on very different timescales, so we can use the adiabatic approximation to reduce the system to a single degree of freedom. The ``secular'' Hamiltonian function writes finally:
   \begin{equation}\label{eq:Fsec}
      \mathcal{F}\big(J,U,V,u\big) = \mathcal{K}(\Sigma_0,U,V,\sigma_0,u)
   \end{equation}
   where $(\Sigma_0,\sigma_0)$ represents any point of the level curve of $\mathcal{K}$ for $(U,u)$ fixed which encloses the signed area $2\pi J$. The momentum $J$ is a secular constant of motion, related to the oscillation amplitude of $(\Sigma,\sigma)$. It is negative if the particle evolves inside the resonance island. Since the secular system has only one degree of freedom, the dynamics can be described by the level curves of~\eqref{eq:Fsec} in the plane $(U,u)$ with $V$ and $J$ as parameters.
   
   The interpretation of these level curves is helped by the choice of a ``reference semi-major axis'' $a_0$ for the resonance considered. Indeed, the constant $V$ can be replaced by:
   \begin{equation}
      \eta_0 = V/\sqrt{\mu a_0} + k_p/k = \sqrt{1-\tilde{e}^2}\cos\tilde{I}
   \end{equation}
   where $\tilde{e}$ and $\tilde{I}$ are called the ``reference'' eccentricity and inclination. In simple words, these variables are equal to the secular elements of the particle when its semi-major axis is equal to $a_0$. Then, the variable $U$ is equivalent to the reference perihelion distance $\tilde{q}=a_0(1-\tilde{e})$. The level curves of~\eqref{eq:Fsec} can be plotted in the plane $(\tilde{q},\omega)$ which is much more self-explanatory than its canonical counterpart $(U,u)$.
   
   In the following, this model is applied considering the four giant planets of the Solar System, the mass of the inner ones being added to the Sun. The secular parameters $(\eta_0,J)$ of the objects studied in this paper are computed using the method described in the following section. They are gathered in Tab.~\ref{tab:list}, along with the constants $a_0$ chosen. The current positions $(\omega,\tilde{q})$ of the objects are also given.
   
   \subsection{Determination of the secular parameters}\label{ssec:param}
   In the previous section, we described the changes of coordinates leading to the resonant secular model. The problem is to compute this set of coordinates for a given object, for which only the osculating coordinates are known. The standard procedure consists in digitally filtering the output of a medium-term numerical integration, thus removing the short-period component of the trajectory as we did in the semi-analytical theory. The detailed procedure was introduced by \citet{CARPINO-etal_1987} and implemented in the OrbFit Software Package\footnote{http://adams.dm.unipi.it/orbfit/}. For all the objects studied in this article, a $500\,000$-year numerical integration was found more than enough to cover the semi-secular oscillations (related to the resonant angle). The numerical integrations were made using the same software package, taking the nominal orbits given by AstDyS database\footnote{http://hamilton.dm.unipi.it/astdys/} as initial conditions. The four giant planets were integrated consistently and the masses of the inner ones were added to the Sun. The invariable plane of the Solar System was taken as reference plane for the output. We used heliocentric coordinates, as this is the system on which is based the semi-analytical theory\footnote{Since the temporal series obtained are meant to be filtered, the short-period oscillations of the barycentre of the Solar System, usually removed by using barycentric coordinates, are not a problem.}.
   
\begin{figure}
      \centering
      \includegraphics[width=0.9\hsize]{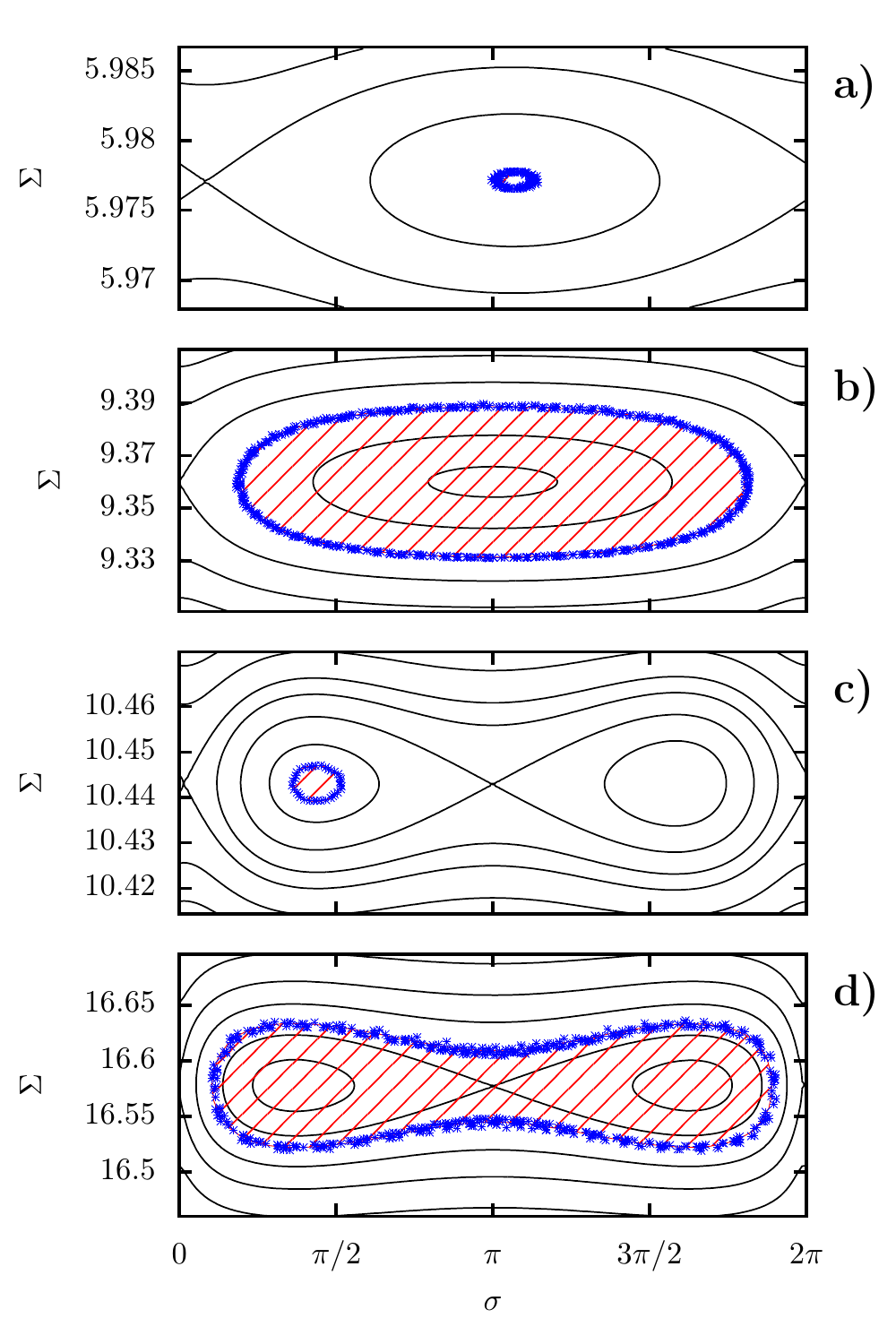}
      \caption{Computation of the secular constant of motion $J$ from the filtered numerical integration. The examples shown are (from top to bottom): $82075$, $119068$, 2008\,ST$_{291}$ and $136120$. See Tab~\ref{tab:list} for the corresponding resonances. The axes are the resonant angle $\sigma$ from Eq.~\ref{eq:sigma} (rad), and its conjugated momentum $\Sigma$ from Eq.~\ref{eq:SUV} (au$^2$rad/yr). The blue crosses come from the filtered output, whereas the red hatched area is equal to the quantity $-2\pi J$ used to construct the secular model. The black lines in the background are the level curves of~\eqref{eq:Kssec} with $(U,u)$ fixed. Various cases are shown: small (\textbf{a},\textbf{c}) or large (\textbf{b},\textbf{d}) area, single (\textbf{a},\textbf{b}) or double (\textbf{c},\textbf{d}) resonance island, simple (\textbf{a},\textbf{b}), asymmetric (\textbf{c}) or horseshoe (\textbf{d}) oscillations.}
      \label{fig:SemiSec}
   \end{figure}

   The secular values of $(U,V,u)$ can be directly picked up from the filtered series, whereas the parameter $J$ requires the computation of the area enclosed by the trajectory in the $(\Sigma,\sigma)$ plane. Fig.~\ref{fig:SemiSec} shows some examples of filtered output, with the graphical representation of the corresponding areas $2\pi J$. In the background, the level curves of~\eqref{eq:Kssec} with $(U,u)$ fixed at their secular (that is filtered) current values are also plotted in order to assess the validity of the semi-analytical procedure. The filtered trajectories were always found to follow pretty well the level curves, showing that our simplified and averaged model captures the essence of the dynamics. The characteristics of the corresponding secular models are gathered in Tab.~\ref{tab:list}. The oscillation amplitudes of the resonant angles were generally found to be in very good agreement with the values given in \cite{LYKAWKA-MUKAI_2007} and \cite{GLADMAN-etal_2012}.
   
   In our numerical integrations, some objects were found outside the expected resonances (blank lines in Tab.~\ref{tab:list}). Apart from two exceptions, this concerns small bodies reported by \citet{SHEPPARD-etal_2016} which are not classified as ``resonant'' from their current orbit but from their probable dynamical history. For these objects, a secular model near the resonance could still be developed (in which $2\pi J$ is the area under the curve), or the uncertainties of the nominal orbit could be taken into account to develop a secular model for a ``potential'' resonant orbit. This will be realised for 2014\,FZ$_{71}$.

\section{General results}\label{sec:res}
   \subsection{Typical resonant secular evolutions}\label{ssec:gen}
\begin{figure*}
      \centering
      \includegraphics[width=0.8\hsize]{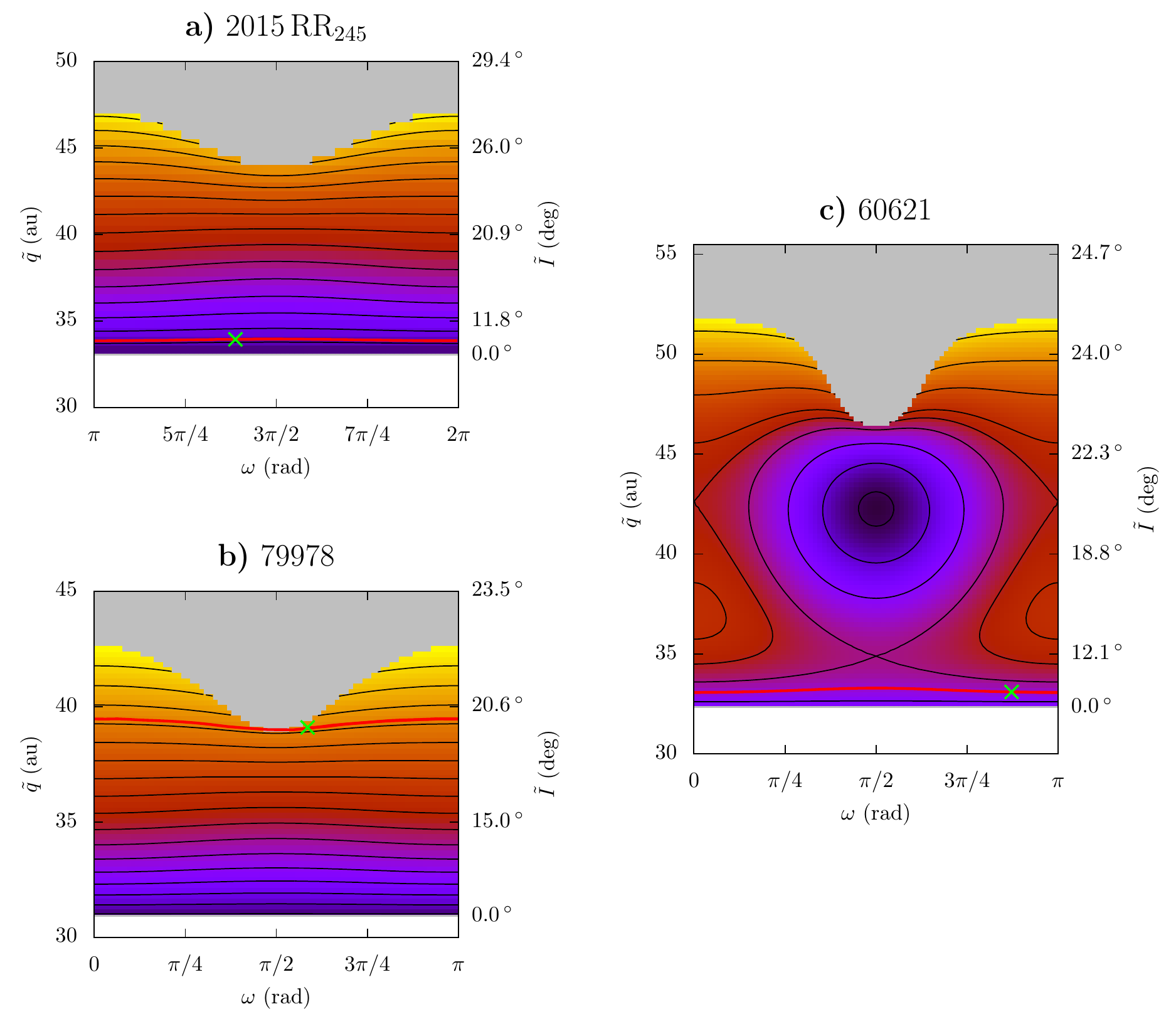}
      \caption{Typical examples of phase portraits for the objects studied in this paper. The colour shades represent the value of the secular Hamiltonian (dark/light for low/high), from which can be deduced the direction of motion along the level curves (black lines). The white zones are forbidden by the value of the parameter $\eta_0$ (it would require a cosine of inclination larger than $1$), and the grey zones are forbidden by the value of the parameter $J$ (the resonance island is too narrow to contain the signed area $2\pi J$). The green cross represents the current position of the object, and the red curve comes from a numerical integration of the two-degree-of-freedom semi-secular system. The names are written above the graphs (see Tab.~\ref{tab:list} for the parameters of the models). On the graph \textbf{b}, the secular trajectory of 79978 forces periodically $\sigma$ to circulate when the curve enters the grey zone (separatrix crossing). The numerical integration shows that this does not significantly affect the overall dynamics (red curve). The graph \textbf{c} shows that 60621 follows a ``flat'' trajectory even if the parameters of the model could allow interesting variations of orbital elements. The behaviour of the object 2004\,EG$_{96}$ (not shown) is very similar to that of 60621, but it follows a level curve much closer to the secular separatrix (lying just above the red curve). A small diffusion of its orbital elements (or a refinement of its orbital solution) could thus put it on a level curve leading to much higher perihelion distances.}
      \label{fig:typical}
   \end{figure*}

   Most of the objects studied in this paper are not very affected by their resonant relation with Neptune: this results in a ``flat'' secular evolution, with circulating $\omega$ and almost constant $\tilde{q}$. Such a secular behaviour is similar to what we would obtain for a non-resonant dynamics (where the only notable features are located at high inclinations). Some examples of level curves of the secular Hamiltonian~\eqref{eq:Fsec} are shown in Fig.~\ref{fig:typical}. The graphs are $\pi$-periodic in $\omega$. The vast majority of objects studied in this paper follow these typical secular trajectories, with the exception of the objects 135571, 2004\,KZ$_{18}$, 2008\,ST$_{291}$, 82075, 2015\,FJ$_{345}$ and 2014\,FZ$_{71}$, which we describe in Sects.~\ref{ssec:libr} and \ref{ssec:FJ345}. In order to check the validity of the adiabatic approximation, each graph also features a numerical integration of the two-degree-of-freedom semi-secular system. As required, the mean trajectory follows a level curve of the secular Hamiltonian. The extra oscillations due to the second degree of freedom are even almost undistinguishable, hidden in the curve width. Fig.~\ref{fig:typical}\textbf{a} presents the most common case: the parameters of the secular model do not allow any peculiar geometry, and the specific level curve followed by the particle has nothing special either. In Fig.~\ref{fig:typical}\textbf{b}, the particle is pushed outside of the resonance island in some portions of its trajectory. This does not affect the overall secular dynamics though, since it re-enters the resonance with a similar parameter $J$. Finally, Fig.~\ref{fig:typical}\textbf{c} presents a ``missed'' interesting case: the parameters of the secular model do allow important variations of the orbital elements but the particle is located on a flat level curve, outside of any libration island.
   
   \subsection{Objects affected by secular libration islands}\label{ssec:libr}
\begin{figure*}
      \centering
      \includegraphics[width=0.8\hsize]{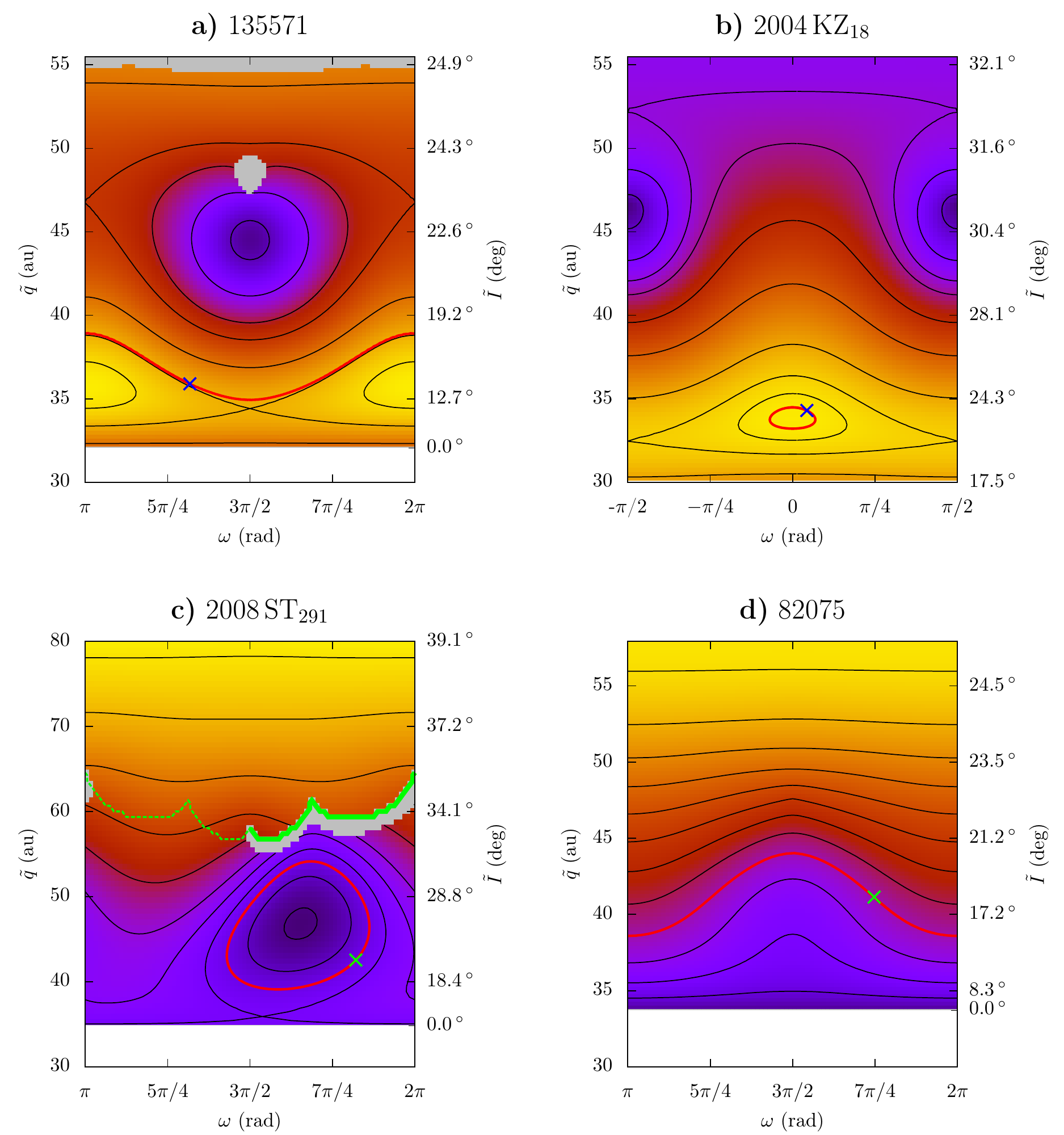}
      \caption{Phase portraits of the objects affected by secular libration islands. On the graph \textbf{a}, the object 135571 is the only one found to have a retrograde circulation of $\omega$. On the graph \textbf{b}, the secular evolution of 2004\,KZ$_{18}$ inside the mean-motion resonance makes $\omega$ oscillate around $0$. On the graph \textbf{c}, there are two distinct regions according to the topology of the semi-secular phase-space, since the resonance of 2008\,ST$_{291}$ is of type $1\!:\!k$. Below the green line, the resonance island is doubled (as in Fig.~\ref{fig:SemiSec}\textbf{c},\textbf{d}) resulting in asymmetric level curves. Above the green line, only one resonance island remains (as in Fig.~\ref{fig:SemiSec}\textbf{a},\textbf{b}) and the level curves are symmetric. If the line is crossed, there can be either a discontinuity (thick green line) if the particle is located in the vanishing island, or a smooth transition (dotted green line) if it is located in the persisting island.}
      \label{fig:libisl}
   \end{figure*}

   The object 135571, presented in Fig.~\ref{fig:libisl}\textbf{a}, is the only one showing a retrograde circulation of $\omega$. Indeed, it follows a level curve located above the two maxima of the secular Hamiltonian~\eqref{eq:Fsec}. It is also very close to the separatrix: a small chaotic diffusion (or a refinement of the orbital solution) could easily put it inside the secular libration island. The object 2004\,KZ$_{18}$, on the other hand, is clearly located in the island (Fig.~\ref{fig:libisl}\textbf{b}). This smooth quasi-periodic trajectory may not be ``permanent'', however, since it involves small perihelion distances. Indeed, the neighbouring resonances have a large width and a chaotic drift out of the resonance is a serious risk.
   
   Figure~\ref{fig:libisl}\textbf{c} shows the case of 2008\,ST$_{291}$ which is associated with a much more distant resonance. The resonant angle $\sigma$ oscillates in an asymmetric libration island (Fig.~\ref{fig:SemiSec}\textbf{c}), which results in a secular Hamiltonian with level curves that are also asymmetric. 2008\,ST$_{291}$ is located inside a secular libration island, confining $\omega$ around a fixed value and producing large-amplitude oscillations of the perihelion distance. The object is on the descending part of the trajectory, implying that it has completed at least one cycle (period of about $115$~Myrs). However, we note that 2008\,ST$_{291}$ has still uncertain orbital elements: \cite{SHEPPARD-etal_2016} reported clones evolving from $q\approx 35$ to $60$~au. These limits can be roughly measured in Fig.~\ref{fig:libisl}\textbf{c}, according to the position of the separatrix with respect to the object.
   
   Finally we can also mention the object 82075, presented in Fig.~\ref{fig:libisl}\textbf{d}, which does not interact directly with a secular libration island but which is located on a level curve producing rather large variations of orbital elements (circulation of $\omega$ and oscillations of $\tilde{q}$ from about $38$ to $44$~au).
   
   \subsection{Regular-by-part secular dynamics}\label{ssec:FJ345}
\begin{figure*}
      \centering
      \includegraphics[width=0.83\hsize]{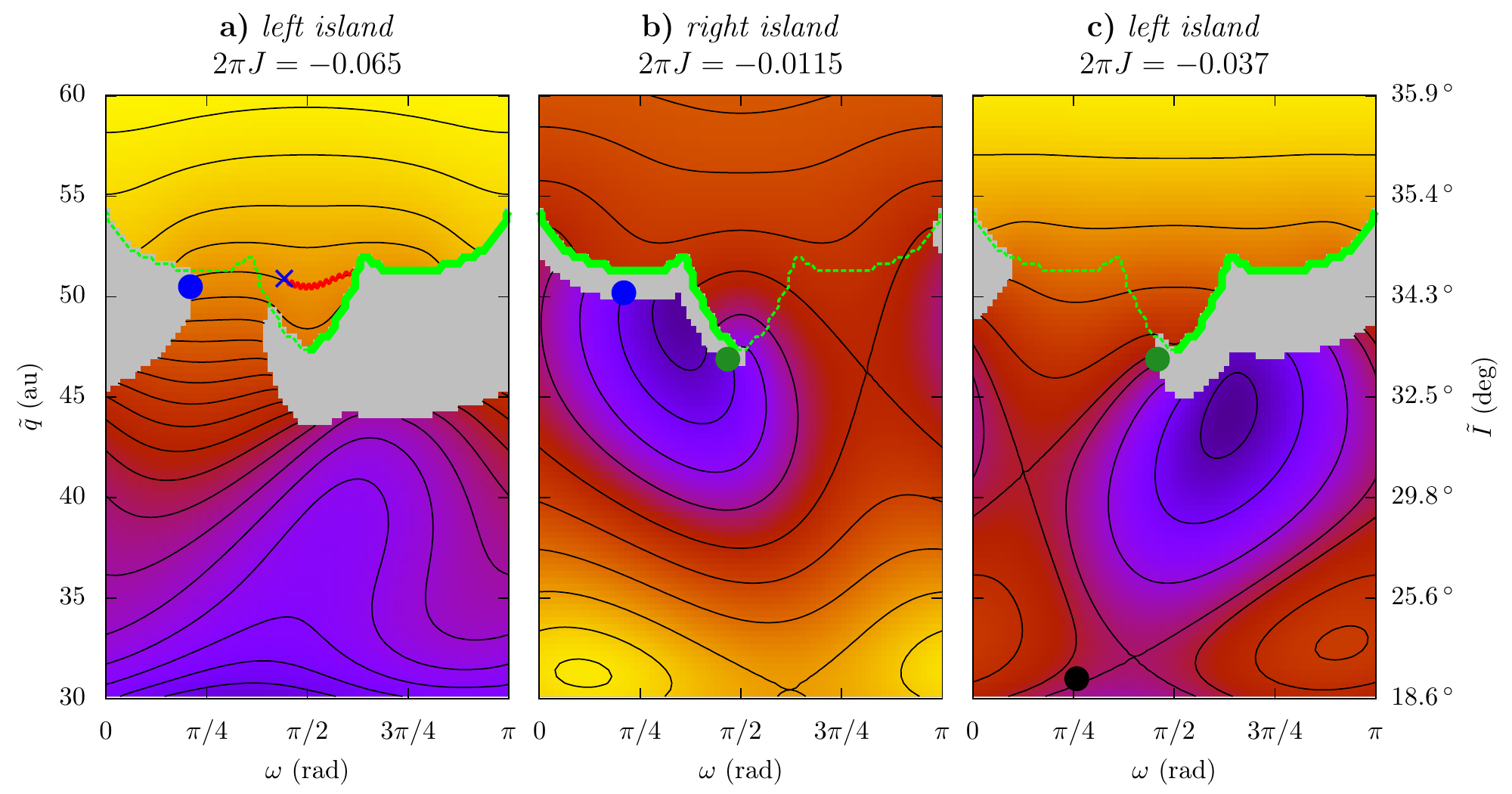}
      \caption{Level curves of the secular Hamiltonian for the current and past evolution of the object 2015\,FJ$_{345}$. Three smooth trajectories are considered, divided by separatrix crossings. The constant parameter $\eta_0$ is given in Tab.~\ref{tab:list}, whereas $2\pi J$ is given above the graphs (au$^2$rad$^2$/yr). The current orbital elements of 2015\,FJ$_{345}$ result in the leftmost phase portrait \textbf{a}. Contrary to previous graphs, the secondary oscillations of the semi-secular trajectory (red curve) are noticeable. The secular trajectory (mean curve), though, follows still pretty well the level curves given by the model. Starting from the current position of 2015\,FJ$_{345}$ (blue cross), the red curve is drawn until it reaches a change of the resonance topology (from a single to a double island), on the green line. Following the level curve toward the left (\emph{past} evolution), the blue spot represents the position of the previous separatrix crossing undergone by 2015\,FJ$_{345}$. The middle graph \textbf{b} shows the secular model for the past evolution of 2015\,FJ$_{345}$. The blue spot represents its position at the transition to its current state (same as graph \textbf{a}). Following the level curve downwards (past evolution), the green spot shows the position of the previous separatrix crossing undergone by 2015\,FJ$_{345}$. The rightmost graph \textbf{c} presents the secular model for the earlier evolution of 2015\,FJ$_{345}$. The green spot shows its position at the transition to the next part of the trajectory (same as graph \textbf{b}). Following the level curve downwards (past evolution), the black spot is taken as initial condition for a numerical integration (see text and Fig.~\ref{fig:2015FJ345_SS}).}
      \label{fig:2015FJ345_all}
   \end{figure*}

   The object 2015\,FJ$_{345}$ presents the most interesting secular dynamics. Figure~\ref{fig:2015FJ345_all}\textbf{a} shows that it is currently located in the single-island region of the $1\!:\!3$ resonance (that is above the ``green line''), but on a level curve coming from and leading to discontinuous transitions. Hence, its perihelion distance was probably raised from the double-island region (below the green line), starting much closer to Neptune, on a secular trajectory leading to the transition. The resonant secular model can be used to reconstruct the detailed scenario: following the secular level curve of 2015\,FJ$_{345}$ backwards (toward the left), its intersection with the grey zone gives its location when it was ejected from the vanishing resonance island. At this point, the total area available in this island gives the parameter $J$ to be used in a secular model describing the previous portion of the secular dynamics. Proceeding this way, we suppose that on a secular timescale, the separatrix crossings can be considered as instantaneous. A chaotic extra change of $J$ should be expected in the real case (depending on the exact phase of $\Sigma$ and $\sigma$ at the moment of the crossing), but it is expected to be negligible if the adiabatic approximation is well verified. The area measured is $2\pi J\approx -0.0115$ au$^2$rad$^2$/yr, which is quite small since the separatrix crossing happened near the green line (blue spot on Fig.~\ref{fig:2015FJ345_all}\textbf{a}). The corresponding previous secular trajectory is shown in Fig.~\ref{fig:2015FJ345_all}\textbf{b}. Surprisingly, we find once again a trajectory coming from and leading to a transition, without any passage to smaller perihelion distances, where the resonance capture with Neptune could have occurred. We use the same method to get the secular trajectory before this second transition (green spot in Fig.~\ref{fig:2015FJ345_all}\textbf{b}). The level curves of the corresponding secular Hamiltonian are presented in Fig.~\ref{fig:2015FJ345_all}\textbf{c}: we finally get a secular trajectory leading to much lower perihelion distances (at least below $\tilde{q}=35$~au, depending on the exact level curve followed by the particle), where the capture in resonance with Neptune probably occurred.
   
\begin{figure*}
      \centering
      \includegraphics[width=\hsize]{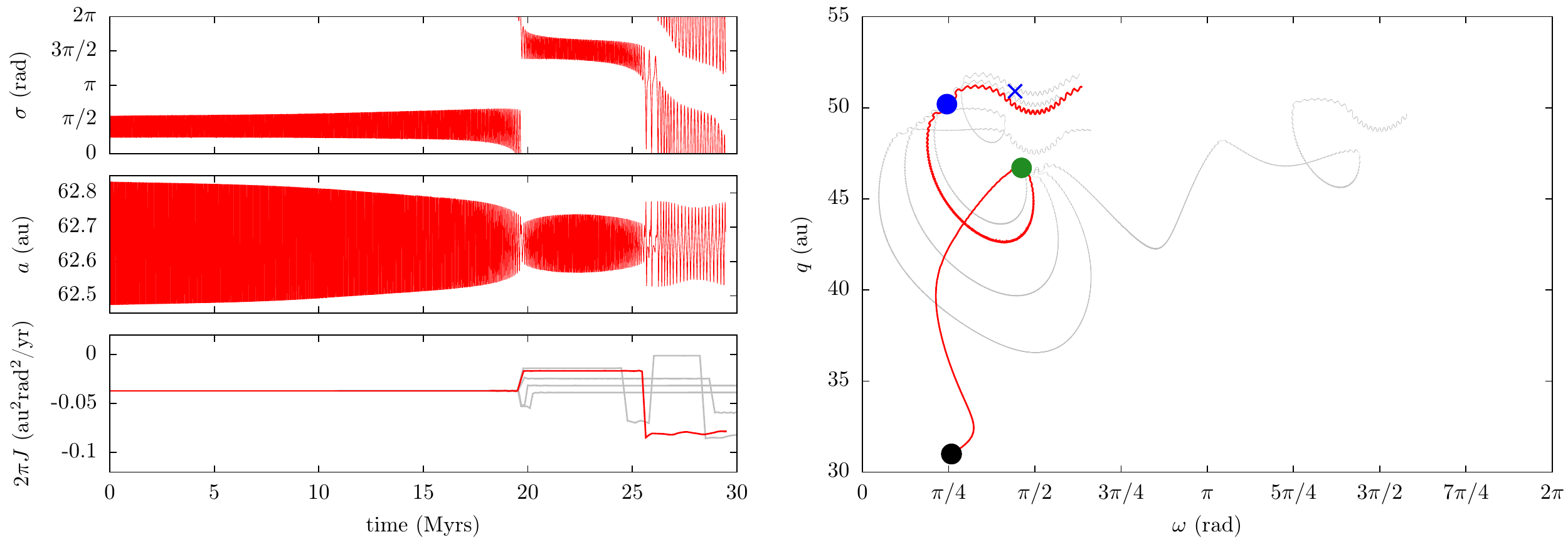}
      \caption{Numerical integration of the two-degree-of-freedom semi-secular system. The initial conditions (origin of the time and black spot on the right) are taken according to the level curves plotted in Fig.~\ref{fig:2015FJ345_all}\textbf{c}. On the right, the evolution of the couple $(U,u)$ is drawn (with $q$ instead of $U$, obtained by inverting Eq.~\ref{eq:SUV}). Several initial phases for $(\sigma,\Sigma)$ were tried, leading to secular trajectories diverging at the first separatrix crossing (grey curves). Among them, the red trajectory is in agreement with our backward reconstruction: its second transition occurs on the blue spot, and it passes near the current position of 2015\,FJ$_{345}$ (to be compared with the level curves from Fig.~\ref{fig:2015FJ345_all}). On the left, the evolution of the couple $(\sigma,\Sigma)$ is represented for the red trajectory (with $a$ instead of $\Sigma$). The two separatrix crossings, leading to a change of oscillation island, are easily noticeable. The bottom graph presents the evolution of the area $2\pi J$ for all the trajectories plotted on the right, showing the divergence at the first transition. Even if $2\pi J$ is plotted as a continuous line, remember that its definition changes after each transition (different resonance island).}
      \label{fig:2015FJ345_SS}
   \end{figure*}

   Since the overall trajectory is rather complicated, Fig.~\ref{fig:2015FJ345_SS} sums up the three components by showing a numerical integration of the two-degree-of-freedom semi-secular system. We started from a position $(\omega,\tilde{q})$ on the level curve of Fig.~\ref{fig:2015FJ345_all}\textbf{c}, as a potential position of capture in resonance with Neptune (black spot). The two separatrix crossings were found to be a bit sensitive to the initial phase chosen for $\sigma$ and $\Sigma$, so we tried different values distributed all along the level curve of the semi-secular Hamiltonian. The grey trajectories in Fig.~\ref{fig:2015FJ345_SS} give an idea of the range of possible behaviours. Roughly half of the integrated trajectories were found to follow qualitatively the scenario from Fig.~\ref{fig:2015FJ345_all} (the red curve is one example of them). As predicted, the trajectory ends up near the observed position of 2015\,FJ$_{345}$, showing that this scenario is dynamically possible. This also gives the timescale involved, which counts in Myrs. In the future, 2015\,FJ$_{345}$ is expected to go on with these types of transitions, possibly turning back to small perihelion distances. In particular, it is possible that several loops already occurred between its capture in resonance and its observed position. It could also stay in a resonant high-perihelion state for Gyrs if it has triggered a ``high-perihelion trapping mechanism'' \citep{SAILLENFEST-etal_2016b}. The orbital uncertainties, though, prevent from any definitive conclusion.
   
\begin{figure}
      \centering
      \includegraphics[width=0.8\hsize]{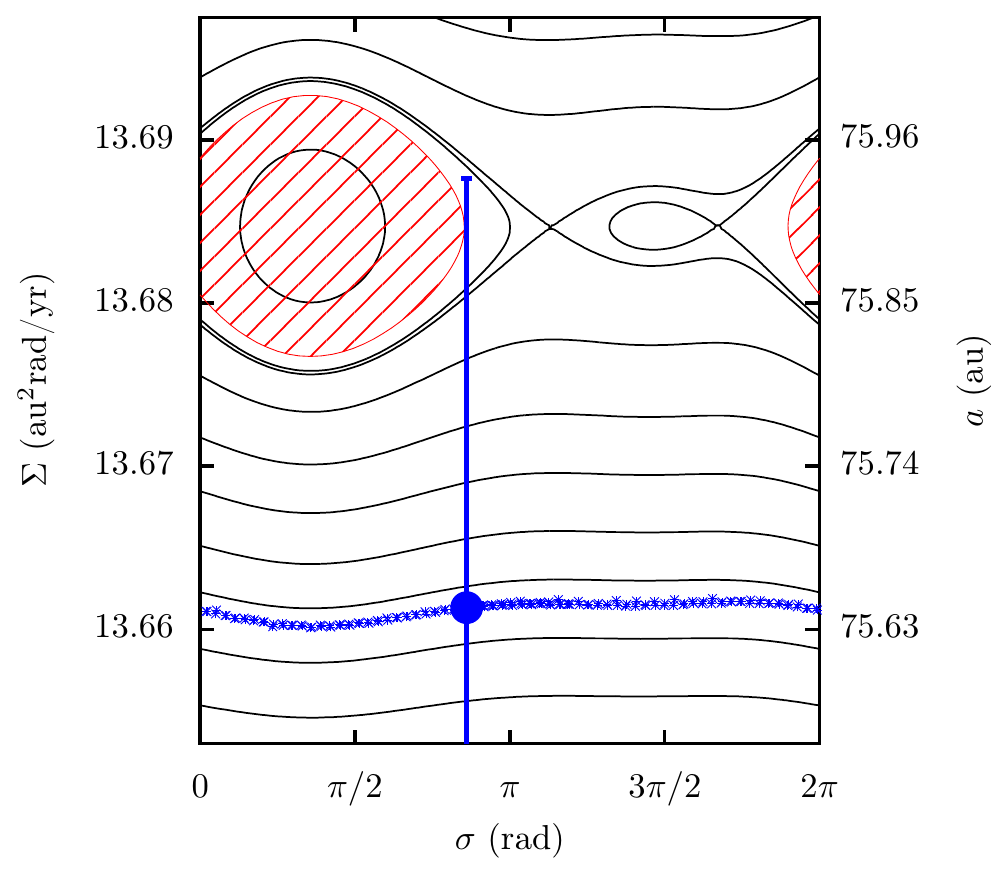}
      \caption{Semi-secular phase portrait of 2014\,FZ$_{71}$. On the right axis, the values of $a$ corresponding to $\Sigma$ are shown. The blue crosses come from the filtered output of the full non-averaged numerical integration (same as Fig.~\ref{fig:SemiSec}), with the current position of 2014\,FZ$_{71}$ on the big spot. For comparison, the 1-sigma error bar of the (osculating) orbital solution is added. The uncertainty is compatible with the red hatched area inside the resonance, from which we get a possible value of the parameter $J$. The same method was used in \citet{SAILLENFEST-etal_2016b}, but it was not restricted to the uncertainty ranges.}
      \label{fig:SemiSec_2014FZ71}
   \end{figure}

\begin{figure*}
      \centering
      \includegraphics[width=0.7\hsize]{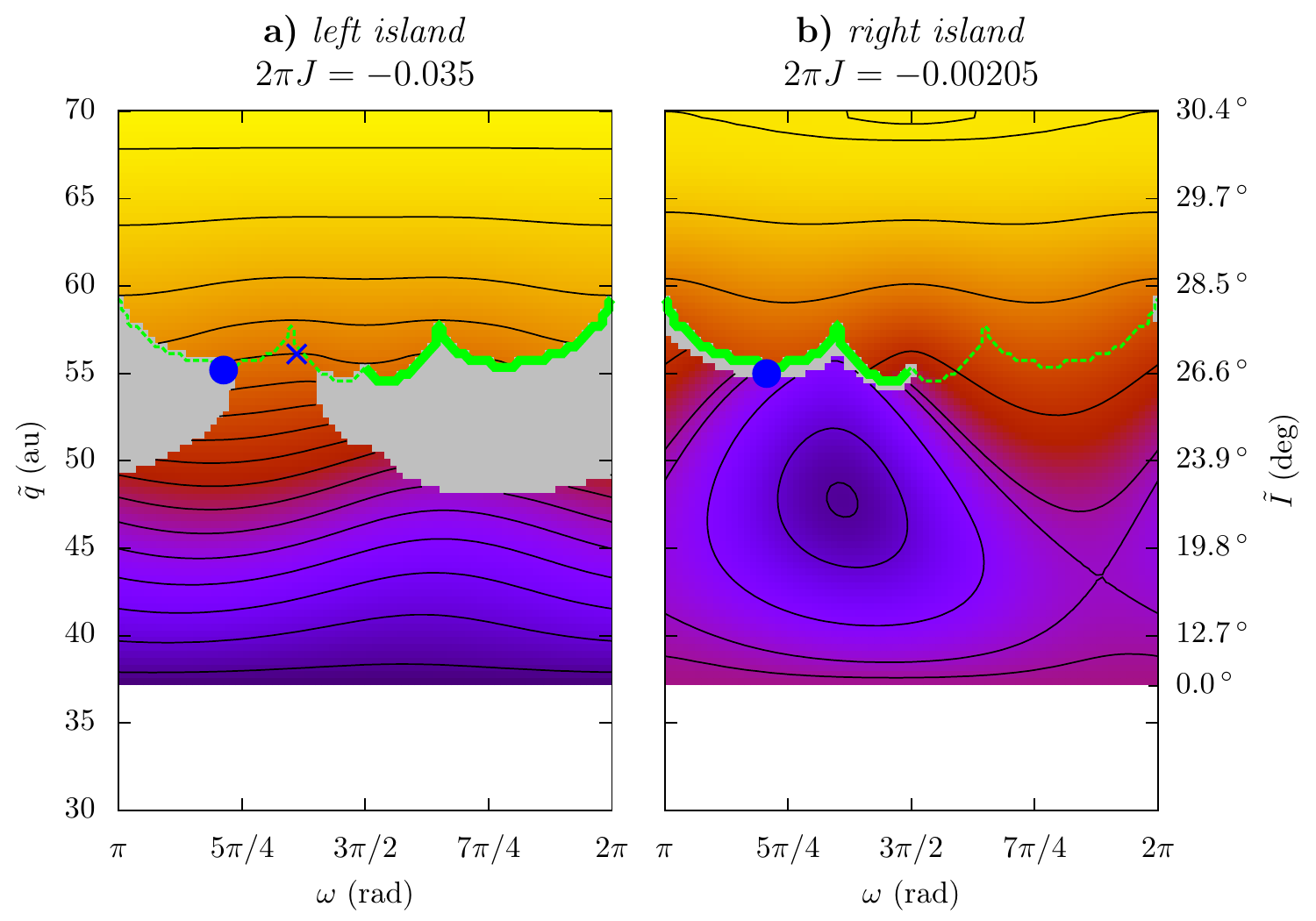}
      \caption{Level curves of the secular Hamiltonian for the current and past evolution of the object 2014\,FZ$_{71}$. Two smooth trajectories are considered, divided by a separatrix crossing. The constant parameter $\eta_0$ is given in Tab.~\ref{tab:list}, whereas $2\pi J$ is given above the graphs (au$^2$rad$^2$/yr). The current orbital elements of 2014\,FZ$_{71}$ result in the left phase portrait \textbf{a}. Following the level curve toward the left (\emph{past} evolution), the blue spot shows the position of the previous separatrix crossing underwent by 2014\,FZ$_{71}$. The right graph \textbf{b} presents the secular model for the past evolution of 2014\,FZ$_{71}$. The blue spot shows its position at the transition to its current state (same as graph \textbf{a}).}
      \label{fig:2014FZ71_all}
   \end{figure*}

   As discussed by \citet{SHEPPARD-etal_2016}, such a mechanism is probably also responsible for the high perihelion distance of 2014\,FZ$_{71}$ even if this object is not found currently in resonance (according to its nominal orbit). Consequently, it could have been left in its current position at the end of Neptune's migration \citep{GOMES-etal_2005,GOMES-etal_2008}, or its resonant secular dynamics itself could have pushed it out of the resonance separatrices. We studied this last scenario by a close-to-resonance secular model \citep{SAILLENFEST-etal_2016a}, but no interaction at all with the resonance was detected (flat level curves). The migration of Neptune set apart, 2014\,FZ$_{71}$ is thus probably much closer to the resonance than its current best-fit orbit seems to indicate. Actually, the uncertainties are even compatible with a trapping in resonance \citep[as already reported by][]{SHEPPARD-etal_2016}. As shown in Fig.~\ref{fig:SemiSec_2014FZ71}, the semi-major axis of 2014\,FZ$_{71}$ can be slightly modified (at the 1-sigma level) to enter the resonance. If we change only the value of $a$, the minimum area $2\pi J$ in the case of resonance is equal to $-0.035$ au$^2$rad$^2$/yr (red hatched region in Fig.~\ref{fig:SemiSec_2014FZ71}). The corresponding resonant secular model is presented in Fig.~\ref{fig:2014FZ71_all}\textbf{a}: we get a geometry very similar to what we obtained for 2015\,FJ$_{345}$, indicating that an analogous long-term resonant evolution probably occurred. Following the level curve toward the past, the second resonance island has a total area of $-0.00205$ au$^2$rad$^2$/yr on the transition line, which gives the secular model represented in Fig.~\ref{fig:2014FZ71_all}\textbf{b}. This time, we directly obtain a trajectory leading to much smaller perihelion distances ($\sim 38.5$~au), compatible with the numerical experiments by \citet{SHEPPARD-etal_2016}. The parameter $\eta_0$ puts a lower limit on the reachable perihelion distance (independently of the separatrix crossings encountered): this limit is equal to about $37$~au and corresponds to $\tilde{I}=0$. This is quite high for a resonant capture (since a diffusion of $a$ is necessary), but that argument was used by \citet{SHEPPARD-etal_2016} to state that 2014\,FZ$_{71}$ possibly originated in the Scattered Disc.
   
\section{Discussion and conclusion}
   A one-degree-of-freedom secular model was used to study the long-term dynamics of the known trans-Neptunian objects in resonance with Neptune, with semi-major axes larger than $50$~au. This method allows us to visualise their long-term trajectories in the plane $(\omega,q)$, in which the notable features (equilibrium points, libration islands, separatrices) can be located. According to the model, most of them experience virtually no change of perihelion distance, indicating a small influence of the resonance on their long-term dynamics. Indeed, the resonant link can never bring them away from their capture configurations, resulting in unstable transient resonances. This is confirmed for 2015\,RR$_{245}$ by the detailed dynamical study by \citet{BANNISTER-etal_2016}. If an object was left on a distant resonant trajectory by the late migration of Neptune, on the contrary, the constancy of its secular perihelion distance would be a guarantee of stability \citep{GOMES-etal_2008}.
   
   On the other hand, four objects are located near notable features of the phase portraits:
   \begin{itemize}
      \item Locked in the $2\!:\!5$ resonance, $135571$ evolves very close to secular libration islands and shows a retrograde (or possibly oscillatory) evolution of $\omega$.
      \item In the same resonance, 2004\,KZ$_{18}$ is located near the centre of a libration island at $\omega=0$ (a perfect secular evolution would produce very small oscillations of $\omega$ around $0$).
      \item The object $82075$ ($3\!:\!8$ resonance) shows oscillations of the perihelion distance from $38$ to $44$~au (with circulating $\omega$).
      \item In the more distant $1\!:\!6$ resonance, 2008\,ST$_{291}$ evolves in a wide asymmetric libration island centred at $\omega\approx 117^\mathrm{o}$, resulting in oscillations of $q$ from $39$ to $55$~au.
   \end{itemize}
   The resonance captures of $135571$ and 2004\,KZ$_{18}$ probably occurred in the lowest part of their secular trajectories (smallest perihelion distances). However, since their dynamics are quasi-periodic, they will regularly return to their entrance configurations, leading to possible expulsions. On the other hand, the minimum perihelion distances reachable by $82075$ and 2008\,ST$_{291}$ seem a bit high to be considered as capture locations. This led \citet{LYKAWKA-MUKAI_2007} to classify $82075$ as ``detached'', whereas the orbital uncertainties of 2008\,ST$_{291}$ prevent us from reaching a definitive conclusion.
   
   The cases of 2015\,FJ$_{345}$ and 2014\,FZ$_{71}$ are more complicated ($1\!:\!3$ and $1\!:\!4$ resonances). Indeed, they follow secular trajectories leading to separatrix crossings, which makes necessary the use of piecewise secular models. Following the level curves backwards, possible scenarios can be retraced from their capture into resonance to their current positions. They require one or more resonant transitions (passage from one resonance island to the other) and lead to much smaller perihelion distances. However, the separatrix crossings are actually not instantaneous (their exact outcomes depend on the phase at the time of transition), leading to potential trajectories significantly different than the reconstructed ones. This holds especially for these two objects, which present a substantial departure from the adiabatic approximation. Nevertheless, the scenarios presented here are dynamically possible (as shown by numerical integrations): they can be considered as the ones producing the shortest paths from capture into resonance to the observed positions.
   
   The method gives also a precise idea of what type of trajectory can be followed by a given object. This allows us to distinguish which orbits could have been created by a resonant link with Neptune on its current orbit, and which ones have a more complex history (involving the planetary migration or another source of perturbation). The fixed value of $V$, or equivalently the approximate constancy\footnote{Strictly speaking, the quantity $\sqrt{1-e^2}\cos I$ is constant only in a non-resonant secular model. The analogous constant in the resonant case is $V=\sqrt{\mu a} \left(\sqrt{1-e^2}\cos I - k_p/k\right)$, but since the particle is trapped in resonance, $a$ is never far from some given value.} of $\sqrt{1-e^2}\cos I$ used by some authors \citep{GOMES-etal_2008,SHEPPARD-etal_2016}, gives only an upper bound for the variations in perihelion distance (white regions in our figures). Indeed, an object could very well be detached, but with a parameter $V$ allowing in principle any eccentricity and inclination. Consider for instance, a particle located on the highest level curve drawn on Fig.~\ref{fig:2015FJ345_all}\textbf{c}.
   
   We did not try to obtain similar graphs for 2004\,XR$_{190}$ since its location out-of-resonance is securely assessed. Its current position is pretty well explained by an analogous resonant evolution, but occurring during the late migration of Neptune \citep{GOMES_2011}.
   
   Finally, we would like to point out that for all objects except 2014\,FZ$_{71}$, the secular parameters were obtained from the nominal orbital solutions (through the filtered numerical integration). Consequently, the actual parameters could be quite different, especially for the recently discovered objects. The corresponding changes concern mostly the parameter $J$, which is sensitive to the initial conditions. For instance, a slight increase of the area $|2\pi J|$ leads to a wider coverage of the ``grey'' zones (on the sides of which separatrix crossings occur). This should not affect our general conclusions, though: for instance, a slight change of orbital elements would leave 2015\,FJ$_{345}$ in the high-perihelion region and on a trajectory coming from a much lower perihelion distance. If ever it turns out to be currently out of the resonance, the migration of Neptune could also be invoked (same type of trajectory, but in an earlier stage of the Solar System).
   
\begin{acknowledgements}
   This work was partly funded by Paris Sciences et Lettres (PSL). We thank the anonymous referee for her/his detailed comments which led to a much better version of the article.
\end{acknowledgements}
   
\bibliographystyle{aa}
\bibliography{AeA_2017}

\begin{thebibliography}{16}
\expandafter\ifx\csname natexlab\endcsname\relax\def\natexlab#1{#1}\fi

\bibitem[{{Bannister} {et~al.}(2016){Bannister}, {Alexandersen}, {Benecchi},
  {Chen}, {Delsanti}, {Fraser}, {Gladman}, {Granvik}, {Grundy},
  {Guilbert-Lepoutre}, {Gwyn}, {Ip}, {Jakubik}, {Jones}, {Kaib}, {Kavelaars},
  {Lacerda}, {Lawler}, {Lehner}, {Lin}, {Lykawka}, {Marsset}, {Murray-Clay},
  {Noll}, {Parker}, {Petit}, {Pike}, {Rousselot}, {Schwamb}, {Shankman},
  {Veres}, {Vernazza}, {Volk}, {Wang}, \& {Weryk}}]{BANNISTER-etal_2016}
{Bannister}, M.~T., {Alexandersen}, M., {Benecchi}, S.~D., {et~al.} 2016, \aj,
  152, 212

\bibitem[{{Carpino} {et~al.}(1987){Carpino}, {Milani}, \&
  {Nobili}}]{CARPINO-etal_1987}
{Carpino}, M., {Milani}, A., \& {Nobili}, A.~M. 1987, \aap, 181, 182

\bibitem[{{Duncan} {et~al.}(1995){Duncan}, {Levison}, \&
  {Budd}}]{DUNCAN-etal_1995}
{Duncan}, M.~J., {Levison}, H.~F., \& {Budd}, S.~M. 1995, \aj, 110, 3073

\bibitem[{{Gallardo} {et~al.}(2012){Gallardo}, {Hugo}, \&
  {Pais}}]{GALLARDO-etal_2012}
{Gallardo}, T., {Hugo}, G., \& {Pais}, P. 2012, \ica, 220, 392

\bibitem[{{Gladman} {et~al.}(2012){Gladman}, {Lawler}, {Petit}, {Kavelaars},
  {Jones}, {Parker}, {Van Laerhoven}, {Nicholson}, {Rousselot}, {Bieryla}, \&
  {Ashby}}]{GLADMAN-etal_2012}
{Gladman}, B., {Lawler}, S.~M., {Petit}, J.-M., {et~al.} 2012, \aj, 144, 23

\bibitem[{{Gladman} {et~al.}(2008){Gladman}, {Marsden}, \&
  {Vanlaerhoven}}]{GLADMAN-etal_2008}
{Gladman}, B., {Marsden}, B.~G., \& {Vanlaerhoven}, C. 2008, {Nomenclature in
  the Outer Solar System}, ed. M.~A. {Barucci}, H.~{Boehnhardt}, D.~P.
  {Cruikshank}, A.~{Morbidelli}, \& R.~{Dotson}, 43

\bibitem[{{Gomes}(2011)}]{GOMES_2011}
{Gomes}, R.~S. 2011, \ica, 215, 661

\bibitem[{{Gomes} {et~al.}(2008){Gomes}, {Fern Ndez}, {Gallardo}, \&
  {Brunini}}]{GOMES-etal_2008}
{Gomes}, R.~S., {Fern Ndez}, J.~A., {Gallardo}, T., \& {Brunini}, A. 2008, {The
  Scattered Disk: Origins, Dynamics, and End States}, ed. M.~A. {Barucci},
  H.~{Boehnhardt}, D.~P. {Cruikshank}, A.~{Morbidelli}, \& R.~{Dotson}, 259

\bibitem[{{Gomes} {et~al.}(2005){Gomes}, {Gallardo}, {Fern{\'a}ndez}, \&
  {Brunini}}]{GOMES-etal_2005}
{Gomes}, R.~S., {Gallardo}, T., {Fern{\'a}ndez}, J.~A., \& {Brunini}, A. 2005,
  \cmda, 91, 109

\bibitem[{{Knezevic} {et~al.}(1991){Knezevic}, {Milani}, {Farinella},
  {Froeschle}, \& {Froeschle}}]{KNEZEVIC-etal_1991}
{Knezevic}, Z., {Milani}, A., {Farinella}, P., {Froeschle}, C., \& {Froeschle},
  C. 1991, \ica, 93, 316

\bibitem[{{Kozai}(1985)}]{KOZAI_1985}
{Kozai}, Y. 1985, \celmec, 36, 47

\bibitem[{{Lykawka} \& {Mukai}(2007)}]{LYKAWKA-MUKAI_2007}
{Lykawka}, P.~S. \& {Mukai}, T. 2007, \ica, 189, 213

\bibitem[{{Malhotra}(1993)}]{MALHOTRA_1993}
{Malhotra}, R. 1993, \nat, 365, 819

\bibitem[{{Saillenfest} {et~al.}(2016){Saillenfest}, {Fouchard}, {Tommei}, \&
  {Valsecchi}}]{SAILLENFEST-etal_2016a}
{Saillenfest}, M., {Fouchard}, M., {Tommei}, G., \& {Valsecchi}, G.~B. 2016,
  \cmda, 126, 369

\bibitem[{{Saillenfest} {et~al.}(2017){Saillenfest}, {Fouchard}, {Tommei}, \&
  {Valsecchi}}]{SAILLENFEST-etal_2016b}
{Saillenfest}, M., {Fouchard}, M., {Tommei}, G., \& {Valsecchi}, G.~B. 2017,
  \cmda, 127, 477

\bibitem[{{Sheppard} {et~al.}(2016){Sheppard}, {Trujillo}, \&
  {Tholen}}]{SHEPPARD-etal_2016}
{Sheppard}, S.~S., {Trujillo}, C., \& {Tholen}, D.~J. 2016, \apjl, 825, L13

\end{thebibliography}

   \begin{table*}
      \centering
      \begin{tabular}{ll|c||c|c|c|c|c||c|ll|ll}
         Name && $k_p\!:\!k$ & \vertical{LY07} & \vertical{GL08} & \vertical{GL12}   & \vertical{SH16} & \vertical{BA16} & $a_0$ & $\eta_0$ & $-2\pi J$ & $\omega$ & $\tilde{q}$ \\
         \hline
         \hline
                  & 2001\,KG$_{76}$  & $4\!:\! 9$  &$\star$&$\times$&        &        &        & $51.718$ & $0.939$ & $0.02$ & $2.668$ & $34.017$ \\
         \hline
         (42301)  & 2001\,UR$_{163}$ & $4\!:\! 9$  &$\star$&$\times$&        &        &        & $51.719$ & $0.958$ & $0.04$ & $6.148$ & $36.974$ \\
         \hline
         (95625)  & 2002\,GX$_{32}$  & $3\!:\! 7$  &$\star$&$\times$&        &        &        & $52.987$ & $0.901$ & $0.07$ & $3.359$ & $33.146$ \\         
         \hline
                  & 2002\,CZ$_{248}$ & $3\!:\! 7$  &        &        &$\star$&        &        & $52.987$ & $0.919$ & $0.026$ & $5.252$ & $32.413$ \\
         \hline
         (131696) & 2001\,XT$_{254}$ & $3\!:\! 7$  &$\star$&$\times$&$\star$&        &        & $52.988$ & $0.946$ & $0.04$ & $3.294$ & $35.910$ \\
         \hline
         (181867) & 1999\,CV$_{118}$ & $3\!:\! 7$  &        &$\times$&        &        &        & $52.988$ & $0.949$ & $0.04$ & $2.650$ & $37.493$ \\
         \hline
         (79978)  & 1999\,CC$_{158}$ & $5\!:\!12$  &        &$\times$&        &        &        & $53.991$ & $0.904$ & $0.017$ & $1.841$ & $39.096$ \\
         \hline
         (119878) & 2002\,CY$_{224}$ & $5\!:\!12$  &$\star$&$\times$&        &        &        & $53.991$ & $0.896$ & $0.011$ & $2.689$ & $35.275$ \\         
         \hline
         (84522)  & 2002\,TC$_{302}$ & $2\!:\! 5$  &        &$\times$&        &        &        & $55.474$ & $0.784$ & $0.13$ & $1.585$ & $39.002$ \\
         \hline
         (26375)  & 1999\,DE$_{9}$   & $2\!:\! 5$  &$\star$&$\times$&        &        &        & $55.478$ & $0.897$ & $0.18$ & $2.884$ & $32.283$ \\
         \hline
         (60621)  & 2000\,FE$_{8}$   & $2\!:\! 5$  &$\star$&$\times$&$\star$&        &        & $55.479$ & $0.909$ & $0.05$ & $2.739$ & $33.109$ \\
         \hline
         (38084)  & 1999\,HB$_{12}$  & $2\!:\! 5$  &$\star$&$\times$&        &        &        & $55.480$ & $0.890$ & $0.105$ & $1.052$ & $32.615$ \\
         \hline
                  & 2004\,KZ$_{18}$  & $2\!:\! 5$  &        &        &$\star$&        &        & $55.480$ & $0.847$ & $0.0015$ & $0.136$ & $34.310$ \\
         \hline
                  & 2004\,EG$_{96}$  & $2\!:\! 5$  &$\star$&        &$\star$&        &        & $55.481$ & $0.871$ & $0.045$ & $0.185$ & $32.110$ \\
         \hline
                  & 2002\,GP$_{32}$  & $2\!:\! 5$  &$\star$&$\times$&$\star$&        &        & $55.481$ & $0.907$ & $0.0013$ & $0.463$ & $32.081$ \\
         \hline
                  & 2000\,SR$_{331}$ & $2\!:\! 5$  &$\star$&$\times$&        &        &        & $55.481$ & $0.896$ & $0.023$ & $3.812$ & $31.157$ \\
         \hline
         (143707) & 2003\,UY$_{117}$ & $2\!:\! 5$  &$\star$&$\times$&        &        &        & $55.481$ & $0.899$ & $0.064$ & $1.895$ & $32.510$ \\
         \hline
         (135571) & 2002\,GG$_{32}$  & $2\!:\! 5$  &$\star$&$\times$&        &        &        & $55.481$ & $0.907$ & $0.03$ & $4.140$ & $35.905$ \\
         \hline
         (69988)  & 1998\,WA$_{31}$  & $2\!:\! 5$  &$\star$&$\times$&        &        &        & $55.481$ & $0.890$ & $0.12$ & $5.571$ & $31.581$ \\
         \hline
         (119068) & 2001\,KC$_{77}$  & $2\!:\! 5$  &$\star$&$\times$&        &        &        & $55.482$ & $0.913$ & $0.24$ & $3.239$ & $35.455$ \\
         \hline
                  & 2004\,HO$_{79}$  & $2\!:\! 5$  &        &        &$\star$&        &        & $55.482$ & $0.907$ & $0.08$ & $2.359$ & $32.507$ \\
         \hline
                  & 2001\,XQ$_{254}$ & $2\!:\! 5$  &$\star$&$\times$&        &        &        & $55.484$ & $0.894$ & $0.19$ & $0.189$ & $31.089$ \\
         \hline
                  & 2015\,GP$_{50}$  & $2\!:\! 5$  &        &        &        &$\times$&        & & & & & \\
         \hline
                  & 2012\,FH$_{84}$  & $2\!:\! 5$  &        &        &        &$\times$&        & & & & & \\
         \hline
         (82075)  & 2000\,YW$_{134}$ & $3\!:\! 8$  &$\star$&$\times$&        &        &        & $57.920$ & $0.909$ & $0.00035$ & $5.489$ & $41.146$ \\
         \hline
                  & 2004\,XR$_{190}$ & $3\!:\! 8$  &        &        &        &$\times$&        & & & & & \\
         \hline
                  & 2015\,FJ$_{345}$ & $1\!:\! 3$  &        &        &        &$\times$&        & $62.649$ & $0.809$ & $0.065$ & $1.391$ & $50.905$ \\
         \hline
         (136120) & 2003\,LG$_{7}$   & $1\!:\! 3$  &$\star$&$\times$&        &        &        & $62.660$ & $0.817$ & $0.5$ & $5.913$ & $32.483$ \\
         \hline
                  & 2015\,KH$_{162}$ & $1\!:\! 3$  &        &        &        &$\times$&        & & & & & \\
         \hline
                  & 2013\,FQ$_{28}$  & $1\!:\! 3$  &        &        &        &$\times$&        & & & & & \\
         \hline
         (160148) & 2001\,KV$_{76}$  & $2\!:\! 7$  &        &$\times$&        &        &        & $69.430$ & $0.835$ & $0.076$ & $4.098$ & $34.421$ \\
         \hline
         (126619) & 2002\,CX$_{154}$ & $3\!:\!11$  &        &$\times$&        &        &        & & & & & \\
         \hline
                  & 2014\,FC$_{69}$  & $3\!:\!11$  &        &        &        &$\times$&        & & & & & \\
         \hline
                  & 2014\,FZ$_{71}$  & $1\!:\! 4$  &        &        &        &$\times$&        & $75.900$ & $0.860$ & $0.035$ & $4.277$& $56.117$ \\
         \hline
                  & 2015\,RR$_{245}$ & $2\!:\! 9$  &        &        &        &        &$\star$& $82.100$ & $0.802$ & $0.075$ & $4.360$ & $33.942$ \\
         \hline
         (26181)  & 1996\,GQ$_{21}$  & $2\!:\!11$  &        &$\times$&        &        &        & & & & & \\
         \hline
                  & 2008\,ST$_{291}$ & $1\!:\! 6$  &        &        &        &$\times$&        & $99.453$ & $0.761$ & $0.003$ & $5.721$ & $42.566$ \\
         \hline
         (184212) & 2004\,PB$_{112}$ & $5\!:\!27$  &        &$\times$&        &        &        & $107.575$ & $0.713$ & $0.0055$ & $0.158$ & $35.392$
      \end{tabular}
      \vspace{0.3cm}
      \caption{List of the trans-Neptunian objects studied in this paper. The resonance ratios $k_p\!:\!k$ are all with Neptune. The references used are abbreviated in LY07 \citep{LYKAWKA-MUKAI_2007}, GL08 \citep{GLADMAN-etal_2008}, GL12 \citep{GLADMAN-etal_2012}, SH16 \citep{SHEPPARD-etal_2016} and BA16 \citep{BANNISTER-etal_2016}. A cross indicates that the object is classified as ``resonant''; a star indicates that the article provides also the oscillation amplitude of the resonant angle with its uncertainty. In LY07 and GL12, the star is omitted if the classification is said ``insecure''. The right part of the table gives the characteristics of the resonant secular models obtained in this paper: $a_0$ is the reference semi-major axis chosen (au) whereas $\eta_0$ (no unit) and $-2\pi J$ (au$^2$rad$^2$/yr) are the fixed parameters. Finally, $\omega$ (rad) and $\tilde{q}$ (au) are the current secular argument of perihelion and reference perihelion distance of the object (see Sect.~\ref{ssec:model}). Blank fields mean that the object was found non-resonant in the numerical integration starting from the nominal orbit given by AstDyS database. In the particular case of 2014\,FZ$_{71}$, a secular model was developed although the nominal orbital solution was not resonant (see Sect.~\ref{ssec:FJ345}). In theory, the 1-sigma error bars would allow it also for 2015\,GP$_{50}$, 2012\,FH$_{84}$ and 2013\,FQ$_{28}$.}
      \label{tab:list}
   \end{table*}

\end{document}